# Designing for the don't cares: A story about a socio-technical system

Ian Sommerville

## Introduction

This is a story of a software-intensive socio-technical system that I helped design. It's a story rather than an academic paper (although I was an academic for more than 30 years) because it is one person's perception of experiences in designing a complex system. It's opinionated, and I hope, easy to understand. Another team member would write a different story; I don't attempt to present a sanitised collective view on what we did or what was achieved.

If you prefer to read dry academic papers written in the passive voice, hedged with qualifications and with reference lists that nobody reads, I suggest that you stop reading now. This story has none of these things and it will just annoy you.

Like all good stories, this tells a tale of a small team tackling a large problem and succeeding against the odds. It has a happy ending of sorts but, if we can push the analogy a little, it leaves open the door for a sequel where the forces of darkness return to foil the plans of the well-meaning rebels.

The story begins in the summer of 2012 when I had a call from the Chief Scientific Advisor to the Scottish Government. She had a familiar tale to tell of a government IT project that was not going well and asked me if I would be interested in getting involved. The system in question was a digital learning environment (DLE), used by students and teachers, to be installed in all Scottish schools. An existing DLE, called Glow, was installed some years ago but work on a more modern replacement for this had stalled. Stakeholders could not agree on what was required and time was running out before the contract to support the current system ended.

I was asked to be the engineering lead in a small team of educationalists who were tasked with writing the requirements for this new system. This team included teachers, who had used technology to support learning, teenage school students, educational technology experts, and government representatives. I've been interested in complex sociotechnical systems and system requirements for some time and was intrigued by this request, in spite of not knowing anything about digital learning systems. The sting in the tail, however, was the deadline – we needed to finish this work in six months. I thought that this was rather ambitious but, perhaps recklessly, agreed to get involved.

Scotland was probably the first country in the world to deploy a national digital learning environment in 2006. This was an innovative and pioneering system but suffered from two fatal flaws. There were problems with the system's usability and it was a closed system. It offered a set of applications that were designed in the early years of the 21st century. Much more effective and usable applications had become freely available on the Internet before the system was deployed.

Nevertheless, there were a minority of teachers and students who did innovative and exciting things with this system. Students who struggled with writing found that the system helped them create work that they were proud of. They improved their self-esteem and discovered that education was not just a waste of time. Teachers found ways to encourage creativity and students shared their work across the country. Experiments showed that using computer systems to support learning did really work.

Our goal was to specify a replacement for this original system that avoided the flaws in its design and which provided a range of opportunities for teachers to use digital technologies to support learning in their classrooms. We wanted to create an open system that could integrate the web services and applications that teachers and students really wanted to use. Furthermore, we recognised that the days of desktop computers are numbered so teachers, students and their parents had to be able to access this system from anywhere and from any computer or mobile device.

Potentially, this system has a large user base. There are more than 3000 schools in which it will be deployed ranging from tiny rural schools with a handful of pupils and a couple of teachers to huge city school campuses, with thousands of students. There are more than 50,000 teachers and almost potential student users. As an aim was to use the system to involve parents in their childrens' education, there are more than a million potential users of the system.

The key challenges that we faced were:

1. The need to accommodate a range of users from age 3 to (potentially) age 83 . Parents and grandparents were potential system users. An unusual constraint that we had was that some of the most creative users couldn't actually read.

2. The very complex system of governance (which I will try to explain later) for the system involving at least 33 separate bodies.

3. A heterogeneous hardware base, widely differing hardware procurement policies and network access across schools.

4. An operational environment where policies were not necessarily driven by educational considerations but were focused on avoiding reputational and legal risks.

5. A user base that had either never taken up the existing system or who were abandoning its use.

None of these (except maybe the first) are really technical problems and this really brings me to the moral of this story. For many types of user-facing system, the technology is not the problem – it's the socio-technical stuff that causes the real difficulties. Socio-technical factors are the reason why many technically excellent systems don't really work that well. They're the reason why so many software-intensive systems projects are late, over-budget and fail to realise their objectives. As engineers, I think it's time that we faced up to these issues and thought about how we can better understand how they affect the systems that we are creating.

## Background

Our story really starts back in the mists of Internet time in the early years of the 21st century when it was suggested that there should be an 'Intranet' for Scottish schools. The idea of an intranet was fashionable at that time – essentially an internal network for sharing information using Internet protocols. After extensive discussions, the Scottish Government duly let a contract for the development of this system – which came to be called Glow. It went live in 2006 and was probably the first national digital learning environment in the world.

The early years of the 21st century were a time of explosive development for both the web and hardware technologies. When Glow was proposed, there was no such thing as social media or software as a service where applications ran in a browser. Most computers were desktop PCs and the majority of non-corporate internet access was through dial-up connections. By the time it was introduced in 2006, things were starting to change – social media had emerged, Apple was resurgent, laptops were starting to replace desktops and home broadband had arrived. It looked out of date when it was deployed.

Glow was procured by the Scottish Government and local government authorities in Scotland were heavily involved in its governance, as I shall explain later. They therefore applied government security standards to the system. There was central control over security and users were required to change passwords frequently. An implementation decision was made to base the system on a hierarchical file system.

Glow was never widely accepted by the teaching community. Although there were some enthusiastic users who did remarkable things with the system, most teachers found it to be unusable. The main (although not the only) reasons for this were the security policies where students regularly forgot passwords and could not get them immediately reset and the complexity of the hierarchical system. At best, only 10%-15% of the potential user community used Glow and many of them simply used it as a gateway for their email, rather than for innovative approaches to learning.

By 2011, it was clear that Glow as it was, was not fit for its intended purpose. It incorporated a range of special-purpose applications that were far less functional than other freely available systems. Discussions started in the community about a replacement for the system but these were prolonged and inconclusive and it was in 2012 that I became involved as I explained in the introduction.

### Governance

The governance of a system is the mechanism by which policies, regulation and constraints that affect the system are established. It is distinct from management, which is concerned with the implementation of these policies.  Systems are shaped and constrained by the policies that are established by the governance mechanism. If these policies are inappropriate or inconsistent, it means that a system cannot deliver to its full potential. If the governance is complex, then changing policies or establishing a common policy is very difficult and time consuming.

In Scotland, as in many other countries, the provision of education is a shared responsibility of national and local government. National government sets out policies and constraints, provides funding and some shared services. Local government (in Scotland, called local authorities) distribute funding to schools and provide an infrastructure for education including networking, administrative services and hardware. There are 32 local authorities that are legally responsible for providing education. Their principal concern is ensuring that everything that is done in schools in the name of the authority is within the law.

As a digital learning environment connects to both the local authority infrastructure and their administrative systems, they are therefore involved in the governance of the learning environment. The system itself is funded by national government and managed as a national service so a national government body is also involved in the governance of the system.

The governance of the Glow system therefore involves 33 separate bodies. National government does not have the authority to impose its wish on local governments. Furthermore, policies that affect the digital learning environment are not set by a single body but by a number of separate groups within each authority. For example, separate groups exist concerned with child safety, data protection and network provision.

There lack of a coherent system of governance means is that there are few, if any, agreed policies for the system as a whole. Policies are set at a local level and so users of the system in different local authorities actually experience different systems. This is most obviously manifested in Internet filtering policies – in some authorities, teachers can use material on YouTube as part of their lessons, in others only YouTube Education is allowed and in a small number of authorities, YouTube is completely banned. The reasons for these differences stem from different interpretations in different authorities of their child protection responsibilities.

The history and governance of the Glow system profoundly affected the perception of that system by its users:

1. Teachers and students found the existing system unusable and were therefore cynical that any new system procured and managed by the same body would be any better.

2. There was a general perception that those responsible for system governance were more concerned with protecting their reputation than with the provision of an effective educational service.

3. There was a general lack of confidence in the user community in the national body that was responsible for managing and supporting the current DLE. Few of the teachers that I spoke with believed that this body had the vision and competence to deliver a better system.

I've said a fair bit about the background and governance of the current system because its very relevant to our story. These issues have had a profound influence on both the process of trying to understand what should be in a replacement system and in that system's architecture. I'll come back to both of these later in the story.

## Requirements for a digital learning environment

As I have said, we were tasked with creating a specification for a new digital learning environment for Scottish schools. I did a bit of background reading on these systems and, frankly, they didn't seem to be too complex. I reckoned that we could use a methodical approach to deriving the requirements by using and adapting standard requirements engineering methods.

Requirements engineering methods are all much the same. They involve an iterative process that includes:

- Understanding the business need and the environment where the system will be deployed.

- Scoping the system and establishing its boundaries

- Engaging with stakeholders to talk about the requirements for the system. These engagements may be facilitated with models of the existing system or the system that is being proposed.

- Documenting the requirements and checking these, to some extent at least, meet the business need and stakeholder requirements.

I couldn't have been more wrong. In the course of our initial discussions, it was abundantly clear that the problems with this system were political and socio-technical rather than technical.

In this case, there were no real business needs for a system. There was no pressure from teachers for a new system – they were happy to use existing tools outside of a framework and they didn't really believe that technology made much of a difference to learning for most students.  The desire for a new system was, fundamentally, a political one – Scotland had led the way in introducing a national e-learning system in 2006 and politicians were keen to demonstrate that the country was still in the forefront of developments. Indeed, the Secretary of State for Education took a personal interest in the project.

In our initial discussions, it became clear that there would be problems with other activities that are fundamental to requirements engineering methods:

1. Scoping the system. Some people thought the system should simply be a portal to generic tools; others believed that it would be more appropriate to orient the system around some model of pedagogy and yet others thought that the system should be the core of a whole school automation system.

2. Engaging with stakeholders. As I explained in the Background section, the previous system was not well received and there was considerable political interference in its operation. This meant that many stakeholders were disenchanted and unwilling to spend time in engaging with us.

3. System modelling. Whilst engineers are comfortable with abstractions and models, these are actually quite alien to many people, especially if their education is in arts and humanities rather than science. They find it very difficult to relate models to reality.

All of this emerged after a couple of group meetings and I must say that I felt pretty disenchanted with the whole thing.

Should I, at this stage, have simply walked away? I certainly thought about doing this but decided against it because doing so would not have changed the situation one bit. There would still have been political pressure for a system to be developed. The vested commercial interests who provided the existing system would have stepped in and updated their existing system without addressing the real underlying problems. Furthermore, a completely open and flexible system would probably be rejected by local authorities because they could not control access to or use of services. In the event of misuse by students or teachers, the authorities would be legally responsible and their reputation would be damaged. The chances were they would limit what tools and services could be used and accessed by teachers and students.

I thought that we had a responsibility to the education of our children to try to do better than this. And, I enjoy a challenge.

At this stage, about one month into a six-month project, we were faced with the problem of where do we go to from here. Personally, I felt I needed to know more about how the innovative uses of the existing system and other technologies. I needed to talk with teachers and students using technology and see what they were doing. I also felt that it was important to talk with teachers who weren't using the existing system to see if they had any interests in using technologies in education. From these discussions, I hoped that we could start to articulate the requirements for a new system.

Identifying teachers who were innovative users of technology was not that difficult. The members of our group who were teachers all had an interest in this area. They made suggestions based on their own experience and their experiences of talking with colleagues. They knew of interesting projects going on in schools across Scotland. The government officials on the group offered to make arrangements for me and other group members to visit some of these schools and to talk about what they were doing.

Reaching out to the other group was actually more difficult. In fact, the government officials didn't much like the idea of me talking with grumpy teachers. So, I decided to use a potent but rarely discussed weapon in the requirement engineer's armoury – free food. Free food is important because sharing food is a primal human experience and some people are much more likely to talk over lunch than they would in a formal meeting.

My daughter is a history teacher and so I asked her to set up an unofficial meeting with some of her colleagues to discuss the existing system and the use of technology. I arranged to buy this group lunch and we had an informal discussion about their work, what they thought of educational technologies and their views on a replacement system for Glow.

I'll talk about the discussions that we had with teachers in the next part of our story.

## Engaging with stakeholders

A fundamental part of any approach to system requirements engineering is to engage with stakeholders in the system. These engagements help you understand how people work, how they use existing systems, what they think about these systems, what they think is missing from existing systems and the features that they might find useful in a new system. Most stakeholders don't talk directly about 'requirements'. It's the job of the requirements engineer to distil the requirements from the stakeholder discussions, analyses of existing systems, documentation and other information that's available.

However, it's actually quite hard to discuss new systems with the potential users of these systems. A common feature of the systems that I have been involved with over the past 10 or 12 years is that there have been problems in involving potential system users in the requirements process. I think there are three universal reasons for this:

1.  Organisations have worked to make their processes more efficient. This means that people are incredibly busy. If they take time out from their day job to talk about new systems, this means that they end up with a backlog of work to deal with. Consequently, they don't attend general meetings about new systems, don't read long emails or other documents and are reluctant to engage in face-to-face discussions.

2.  By and large, users of existing systems have 'tamed' these systems – they live with their failings and have incorporated these systems into their normal ways of working. They know that new systems will inevitably mean that they have to change their ways of working and repeat the 'domestication' process for the new system. All of this takes time – but their workload will rarely be reduced to compensate for this.

3.  Many users are cynical about the motives for introducing new systems into an organization. This cynicism may come from previous bad experience with systems and a lack of trust in those running the system. This was certainly true for the new DLE. Sometimes, it comes from a distrust of the motives of management who claim to be introducing 'improvements'. Users don't actually see these as improvements to their lives.

What I have found is reflected in the title of this story – most users simply don't care. They don't care whether or not new systems are introduced. They don't care about the features of these systems (they expect any new system to be bad) and they don't care about the schedule or the priorities of the people who are trying to design the new system. Of course, they do care (and complain) when an inappropriate system is introduced but they don't have the time or the inclination to get involved in the development process.

Even for those users who are interested in the system, their schedule means that meetings have to be arranged at their convenience. In my experience, this means that discussions cannot be arranged at short notice so the whole process of engagement is prolonged. In our case, starting in August, we had a deadline to complete work by the end of December. However, some users who were doing inspiring things with the existing system could not be available for discussions before early November.

The school visits that were arranged to discuss the use of technology to support learning were absolutely inspiring. Teachers and students were enthusiastic about technology and were doing wonderful things. Eight year olds were creating their own iBooks, 13 year olds were creating fabulous digital artworks, students were learning the basics of programming by developing their own apps and so on. I came away from these visits completely convinced that what we were doing was fundamentally worthwhile.

One thing I found surprising was how technology could empower students who sometimes felt excluded from the learning process. For example, students who could not write and spell well tended to avoid tasks that involved writing simply because they were ashamed of their own limitations in this area. However, writing on a blog was different – their work looked the same as their peers and spelling checkers sorted out many of their problems. They were far more willing to try and express themselves in this way rather than in the more traditional approaches used in the classroom.

The discussions that we had with teachers who did not use the current system confirmed two of our initial impressions and provided an important additional insight:

1.  The fundamental barriers to use of the existing system were the security policy and the hierarchical model of information organization. The security policy required regular password changes and inevitably, some students in a class forgot their passwords. This required teachers to spend a significant percentage of their time simply getting students connected to the system. As one teacher said to me 'if you have to spend 20 minutes of a 40 minute lesson sorting out password problems then there really isn't much point in using the system'.

    Hierarchical filing systems may seem to be a normal and natural way of organizing information to those of us who have brought up with Unix. However, for a large percentage of the population, they are completely unnatural – just look at how many people have tens or hundreds of icons on their desktops. Users spent too long looking for information in the system and often gave up because of the complexity of the hierarchy.

2.  Teachers did not see the point of spending much time engaging with the specification of a new system. This was partly because they thought there was no need for such a system, although they did understand why politically it might be built. However, given their experience with the current system, they had no confidence that any new system would be useful and usable.

3.  The important new insight that this discussion revealed was that there was no reluctance at all to use technology in education and all of the teachers recognised its merits, as well as its limitations. Some of the teachers were already experimenting with tools outside of the current system. This was an important insight because it meant that the reluctance to use the existing system was due to limitations with that system rather than a more general belief that technology was not effective in supporting education.

At the end of our discussions, it became clear that the idea of creating a comprehensive requirements specification for the Glow+ system was a complete non-starter. There was no general agreement on what tools and services should be part of the system. The huge diversity in needs from teaching very young children who could not read to preparing young adults for jobs or further education meant that there was no point in creating a long list of services that might be included in a system. In fact, from our discussions, we only identified three 'requirements':

1.  *Usability* The Glow+ system must be much easier to use than the existing system.

2.  *Flexibility* Users should be able to use the tools and services that they believed were the most appropriate for the learning involved. There should be no restrictions placed on the use of the system either by local authority Internet filtering policies or by arbitrary choices of what tools should be made available.

3.  *Coolness* The system must look 'cool' or it won't be accepted by young people.

I must say that the requirement for 'coolness' was a new one for me and one that I have no idea how to support it.  Engineers don't have a reputation for being cool and I'm no exception to that.

But it is, in fact, a perfectly valid requirement.  The general issue of how to make a system acceptable to a user community that has the discretion whether or not to use that system is one that has not really been investigated. All too often it is seen as an 'HCI problem' and is ignored in the specification of the system.

The challenge of what to build remained after our interactions with the user community.

## Envisioning a system

Ten weeks into a six-month project, we'd had a small number of meetings with potential end users of the digital learning environment. We had tried to arrange meetings with other stakeholders but, for the reasons I've talked about in the last section, these had not actually taken place. We were three weeks away from a mid-project review with the Secretary of State for Education to discuss what we'd done.

It was clear at this stage that there was no way that we could establish a set of conventional requirements for this system. It was up to us to be creative in illustrating what the system might do and we had to do this in such a way that could be understood by both politicians and civil servants and the broader teaching community.

It was equally obvious that trying to communicate using mock-ups of what the system might look like would not work. The problem with this approach is that readers immediately think the mock-up is an accurate rendition of the system and they then go on to tell you what is wrong with it.  What we wanted was something that could be the basis of constructive engagement so I suggested that we think about developing a set of user stories.

## User stories

User stories were originally developed as part of so-called Extreme Programming (XP), an agile method of software engineering. In that method, a user story was a description of an interaction that a user might have with a system and this description was used to decide on the system features that could be implemented. Of course, this is not really a new idea – scenarios had been used in requirements engineering for some time before they were 'reinvented' in agile methods.

User stories in XP are created by users and are intended to be understandable by those without any technical background. As all other approaches to eliciting requirements for the system had proved to be impractical, I suggested that the group collectively developed a set of user stories that could communicate what the system could do. We were not trying to define requirements but were exploring how the system might be used to help us understand the key features of the system, based on discussions with teachers of the things they wanted to do.

Rather than express stories at the level of detailed interaction, as in the case in agile methods, we let the story developers choose their own level of abstraction. This is an example of a high-level story about how the system (Glow+) might be used in a primary (elementary) school:

> *Jack is a primary school teacher in Ullapool, teaching P6 pupils. He has decided that a class project should be focused around the fishing industry in the area, looking at the history, development and economic impact of fishing. As part of this, pupils are asked to gather and share reminiscences from relatives, use newspaper archives and collect old photographs related to fishing and fishing communities in the area. Pupils use a Glow+ wiki to gather together fishing stories and SCRAN to access newspaper archives and photographs. However, Jack also needs a photo sharing site as he wants pupils to take and comment on each others' photos and to upload scans of old photographs that they may have in their families.*

> *Jack sends an email to a primary school teachers group, which he is a member of to see if anyone can recommend an appropriate system. Two teachers reply and both suggest that he uses KidsTakePics, a photo sharing site that allows teachers to check and moderate content. As KidsTakePics is not integrated with the Glow+ authentication service, he sets up a teacher and a class account. He uses the Glow+ setup service to add KidsTakePics to the services seen by the pupils in his class so that when they log in, they can immediately use the system to upload photos from their phones and class computers.*

Some stories were much more detailed – this is a story about authentication:

> *Emma is a history teacher in the Royal High School in north Edinburgh who works part-time, 4 days per week. She lives in Morningside (south Edinburgh) and has 2 young children who are at the local primary school. She makes a point of always leaving school by 3.30 so that she can pick up her children – therefore, she often works from home, using her own computer.*

*Emma is teaching the history of the First World War and a group of S3 students are visiting the battlefields in northern France. She want to set up a 'battlefields group' where those attending this can share their research about the places they are visiting and share their pictures and thoughts of the visit. From home, she logs onto the Glow+ system using her Facebook credentials. Emma has two accounts in Glow+ – her teacher account and a parent account associated with the local primary school.*

*The system recognises that she is a multiple account owner and asks her to select the account that she wishes to use. She chooses the teacher account and the system generates her personal start up screen. As well as her selected applications, this also shows a number of 'management apps' that help teachers create and manage student groups. Emma selects the 'group wizard' app that recognizes her role and school from her identity information. She presses the 'New Group' button that generates a list of the names of all RHS students. She enters S3 and History in the level and subject boxes and the list is pruned to show only the names of those students who are studying History at the S3 level.*

*More than half of the students are going on the trip so she chooses 'select all' and then goes down the list unticking those students who are not coming on the trip. There are 2 students with the same name on the list so she clicks on the student to see what class group they are in. She uses this information to untick the student who is not coming on the trip. She also adds her teacher colleagues Jamie and Claire to the group. She then gives the group a name and confirms the group creation. This sets up an icon on her Glow+ screen to represent the group, creates an email alias for the group and asks Emma if she wishes to share the group.*

*She shares access to the group with everyone in the group (which means that they also see the icon on their screen) but, to avoid getting too many emails from students restricts sharing of the email alias to Jamie and Claire. Emma then uploads some material from her own laptop on the trip to Glow+ and shares this with the 'Battlefields Group'. This generates an alert to group members that new material is available when they next login to the system. She then uses the flickr app on her screen to log in to flickr – flickr is not integrated with the Glow+ authentication system so she logs in with her own account and creates a private group to share battlefields photos. She uploads some of her own photos from previous visits and posts an invitation to join the group on the Battlefield Groups web page.*

In all, we developed 26 user stories, ranging in length from a single paragraph to 2 pages of text. We made these available to the education community on a wiki and invited them to comment on the stories and to contribute their own. We had many positive comments about how these brought the system to light and a few negative comments about the unreality of some of these scenarios.

User stories in agile software development are also presented as a 'user-friendly' way of expressing requirements. Consequently, I was surprised how difficult some teachers found story development. They found it unnatural to personalise stories and tended to write then in an abstract way. This, I think, reflected both their academic background and the fact that they were technology enthusiasts.

They were so familiar with the use of technology that they found it hard to articulate the details of their interactions.

User stories were a great success. All of the members of the group could relate to them and they were effective in communicating our vision to a wider community. An unexpected bonus of using these stories was that politicians and civil servants with no technical or educational background could relate to them. The Scottish Secretary for Education commented how the stories brought the system to life for him and how refreshing it was to receive a report on a technical topic that he could understand.

The advantage of the user stories for the teaching community was that we made a point of developing stories for all stages of use of the system – from early years to young adults. We did not therefore require teachers to imagine how some scenario might be translated to their needs. I would be exaggerating if I said that we had a lot of responses to these but at least some teachers engaged and gave their opinions. I don't think that this would have been achieved in any other way.

## An architecture for Glow+

By this stage in our story, we have reached mid-November with a deadline to deliver our final report by the end of December. Of course, the end of December actually means mid-December because of Christmas holidays so we had a month left to do something. We had no functional requirements but an understanding that users wanted to use a wide range of tools and services. There was very little point in trying to select from these as, for sure, we couldn't chose features that everybody would want to use.

I therefore suggested that the only way forward was to design a completely configurable system that could be instantiated in different ways for different groups of users. This was built around 4 principles:

1.  Everything should be provided as a service. This means that everything is replaceable and distributable. There are no privileged services. Everything should run in a standard browser.

2.  We would require the implementation of as few services as possible. For political reasons, we needed to have an authentication service so that the system could track, to some extent, who is doing what. We needed a set of configuration services to create different versions of the DLE and, again for political reasons, we needed storage services, to allow for local control.

    Technically, we didn't really need a storage service as services such as Dropbox, Google Drive or Microsoft OneDrive could have been used. However, the legal departments of some local authorities were unwilling to allow this as there were no guarantees that information would be maintained within the EU.

3.  Users could configure their own version of the environment. This configuration would not require detailed technical knowledge. Configuration could be at a local authority level, a school level or an individual teacher level. Of course, a number of 'standard' configurations would be provided for users who did not want to create their own version of the system.

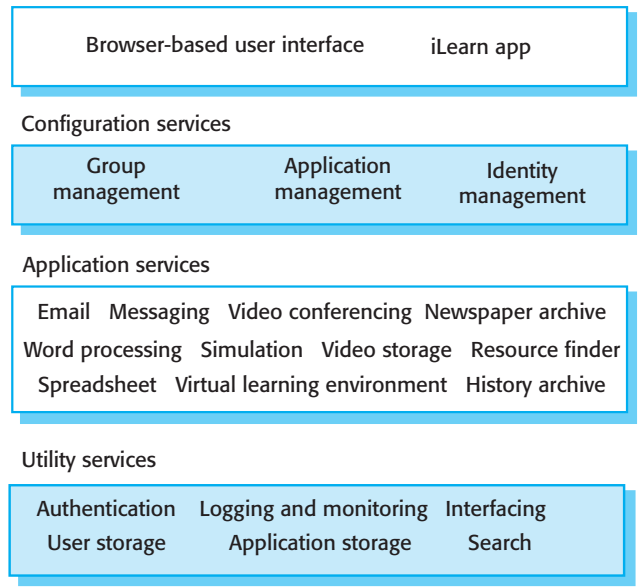

Figure 1. A layered architecture for a DLE

4. Services could be loosely or tightly integrated. Tightly integrated services could use the system's authentication and storage services; loosely integrated services used their own authentication (which could be Google or Facebook authentication) and managed their own storage.

Figure 1 shows that the system has a layered architecture where services are classed as utility services, application services or configuration services. The application services shown here are an example of services that might be made included – this is not a complete set and users can change these as they wish.

The driver for choosing this architecture was the list of challenges that I identified at the start of this story:

1. The need to accommodate a range of users from age 3 to (potentially) age 83. The completely configurable architecture meant that versions of the system could be easily created for different types of user. It was quite possible for these to include different services that did the same thing so that there could be a simple service for people new to the system and a more comprehensive service for those with experience.

2. The very complex system of governance with no single decision making body. While as I would have preferred to change the system governance, we accepted that this was impossible in the time frame available. The flexibility of the system meant that there could be localised versions if necessary, reflecting the policies of each local authority.

3. A heterogeneous hardware base, widely differing hardware procurement policies and network access across schools. By choosing a service-based approach, everything ran in a browser so hardware incompatibilities were minimised. We also recommended that IOS and Android versions of apps to access the system should be made available.

4.  An operational environment where policies were not necessarily driven by educational considerations but were focused on avoiding reputational and legal risks. We proposed an interfacing service so that local authorities could connect the DLE to their own systems that implemented their policies. As I discuss later, we also recommended various actions that we hoped would mean that more enlightened policies could be adopted.

5.  A user base that had either never taken up the existing system or who were abandoning its use. We proposed that the authentication service could make use of Facebook or Google authentication systems so avoiding one of the major critiques of the existing system. The system flexibility meant that teachers and students could use familiar services within the system and we hoped that this would reduce their reluctance to become involved.

The other important benefit of this architecture is that it allowed for incremental delivery and deployment of the new system. A version of the system could be created relatively quickly by using existing authentication services and bundling existing applications and this could be delivered to users to replace the existing system. As more services such as configuration and storage services were developed, these could be deployed without seriously disrupting existing users. We proposed that agile methods of software development should be used with some of the teacher members of the group being involved as proxy users.

Now we're coming to the end of our story and, as I said, it's a happy ending of sorts.

We wrote a report with the user stories and the proposed architecture and presented this to the Government. We raised the problems of governance and recommended that there should be a single set of policies applied to the system rather than policies for each local authority. We emphasised the importance of flexibility and that students and teachers should be able to use the system from their own and school computers.

It was well received and a decision was made to develop the system with a view to deploying it as soon as possible. However, as I'll explain in the next section, it wasn't that easy and the future for the system remains uncertain.

## Lessons learned

Like all large and complex systems, the Glow+ system is unique. It's for a specialised purpose, it has unusually complex governance and it's used in an environment where its users cannot be managed. You might therefore think that there aren't really any general lessons that we can learn from this experience. Well, I think there are, because this kind of system is actually becoming increasingly common as digital services become universal:

1.  More and more systems are being developed for use by professionals to support their work. These professionals have the discretion to accept or reject these systems. Furthermore, these professionals all have access to and experience with generic services available over the Internet and these set a standard that they expect in their professional systems. While this system is specialised for educational professionals, I believe that the lessons learned here are likely to be applicable for systems developed for lawyers, accountants, dentists and other professional users.

2. Complex governance arrangements are becoming increasingly common as new systems are integrated by creating systems of systems, where the constituent systems are owned and managed by different organisations. The increasing power of the media means that more and more organisations are risk averse and anxious to avoid any publicity that affects their reputation.

3. As e-government systems are developed to provide services to citizens, we have a situation where users of the system cannot be managed and there are no sanctions that can be applied for not using the system. Complex security features will alienate users and stop them accessing the system. If the information in these systems is not organised clearly, then people will struggle with their use.

I learned a lot about problems in this project and I don't think that these problems are unique. I've written about these problems because, as far as I can see, few people in the software and systems engineering world publicly acknowledge their existence. As engineers we cannot ignore these issues and focus on technicalities because they have a profound effect on the systems that we are creating.

I haven't written much about answers – I don't have solutions to these problems and I suspect that they have to be handled differently in every system. That's generally the case with socio-technical factors.

I want to end my story with four 'take-away' messages that reflect some of the socio-technical issues that I've discussed here:

1. Methods may be useful where systems are being developed for a clearly defined purpose in a single managed organisation but in more complex organisations, they don't work.

2. Most users don't care about new systems. The key challenges for engineers are finding ways to communicate with these users and building systems that are flexible enough to adapt to different styles of use.

3. Governance is critical – if at all possible, reduce governance complexity as much as you can.

4. User stories work – are by far the best way that I have found to communicate with a wide user community that has diverse backgrounds and experience.

I'll explain these in a bit more detail now.

### Methods

System and software engineering methods always have to make simplifying assumptions. The most common of these assumptions is that these methods are used in a neat, tidy and rational world. Let us call this 'methods world'.

In methods world, system stakeholders are always cooperative and constructive, there are no lawyers raising inconvenient legal issues, managers are committed to the objectives of the organization and procurement and acquisition are never mentioned. Methods world is populated by engineers who speak UML and even understand what strange verb-noun conjunctions such as 'use-case' actually mean. It is a happy, hard-working environment where everyone reads documents and emails, there is no bickering or backstabbing and the whole team work together to build a model of a wondrous system.

Unfortunately, methods world is a mythical place. The real world is not neat and tidy and it certainly isn't rational. System stakeholders pursue their own agendas or, more commonly, are too busy to care. Some stakeholders actively resist the idea of any change whereas others are techies who want to get involved and do the engineer's job for them. Lawyers have more influence than engineers, managers don't turn up for meetings then veto the outcomes and nobody speaks UML. If a model of a system is produced, those who read it (the minority) interpret it according to their prejudices. Those who don't, simply talk about their prejudices.

Of course, the fact that methods don't work that well in practice is old news. I've talked about them here because this story is appearing in a book about the SEMAT method. The SEMAT method was inspired by dissatisfaction with other software engineering methods and was developed to focus on the 'essence of software engineering'.

I think the SEMAT method is well-intentioned and it includes some sound advice. But, sadly, I think it has the same problem as other methods in that it relies on a methods world where all is sweetness and light. Socio-technical issues are not even mentioned in the SEMAT book. Of course, as the authors' say 'this is not the end'. I hope that future developments of the method will start to recognise some of the issues that I've raised here.

### Disengaged users

I believe that disengaged and uninterested users are now the norm for new organisational and e-government systems. Perhaps this has always been the case but I suspect in the early days of computerisation, people were more excited by the technology and could see more clearly how it could improve their work.

Now, there is no real incentive for end users in an organisation to involve themselves in the specification of new systems. They are too busy and, for them, the changes required to adapt to a new way of working are not worth the effort. Of course, they will complain when an unsuitable new system is introduced and may even refuse to use it. But this won't change their mind about participation in the specification process.

I think this has a very important implication for system requirements. 'Flexibility' requirements are relatively uncommon but I believe that flexibility is the key to addressing user disengagement. Instead of defining detailed functionality in advance, we need to have broad functionality areas that are configured and adapted at deployment time to suit different ways of using a system.

The problem here, as we found to some extent, is that procurement policies often require a detailed requirements document so that there can be a 'fair' procurement process. As has been the case with agile methods, both clients and contractors may be reluctant to enter into development contracts with very loosely specified system functionality.

### Governance

Governance complexity was at the root of many of the problems with Glow, the currently used DLE. It led to inappropriate security policies and inconsistent access to services and networks. I suspect that, in spite of our efforts, these are likely to remain issues in a replacement system and this will hinder its adoption and use.

Governance complexity often arises because the different organisations involved in the governance of a system have different perceptions of risks. For example, in this case, a single policy on sharing information was impossible to agree because some authorities considered that there was a legal risk of copyright violation where, for example, teachers used images taken from the Internet in teaching material. Others interpreted the law differently and considered this to be 'fair use' as allowed under copyright laws.

Governance complexity has been largely ignored by the systems engineering community perhaps because most systems are still organisational systems where a single organisation is responsible for the system governance. In the Systems Engineering Body of Knowledge, governance is only discussed at a project level and is conflated with a discussion on project structure.

However, as we create more and more systems of systems, this will become an increasing problem. I believe that it is, perhaps, the major barrier to the effectiveness of large-scale systems of systems.

### User stories

The one unequivocal success in this project was the use of user stories. I remain sceptical about their use in agile software engineering methods where they are used as a substitute for requirements. However, as a means of communicating with and, to some extent, engaging users they were great.

Scenarios, of course, have been around for a long time and have been used in systems engineering to present high-level situation descriptions where systems are used. But, from what I have seen, scenarios are not user stories. They are either very broad descriptions or are simply a narrative description of a system model. It's often hard to disagree with a scenario.

User stories are not this kind of scenario. You can tell a story about how people use the existing system or, as we did, how they might use a new system. But you have to make it personal. That's important because people relate to people and by personalising the story, they can put themselves in the position of the people involved. They can tell you how their story would be the same or what they would do differently.

I suggest that you give them a try.

## What happened next?

Our report was delivered in January 2013 and now, in August 2014, there's still no system. We recommended that agile methods should be used with incremental delivery of the system. While I don't exactly know what's happened over the past 18 months, I know that there have been various factors that have contributed to it:

1. Microsoft offered free use of Office 365 under its educational programme and some people argued that a new system should be built around this suite of services rather than as a loosely integrated system.

2. Systems procurement for public sector systems is governed by European rules that require that an open procurement process should be followed with any European company allowed to bid for this. Procurement staff had no expertise or understanding of agile software development and did not understand how to maintain compliance with procurement regulations.

3. The suppliers of the existing system updated their user interface and argued that this addressed many of Glow's usability problems.

4. After it was finally agreed that the system should be developed internally using an agile approach, there were problems in finding suitably qualified developers to work on the project.

5. The local authorities involved in system governance remain risk averse and insist that various regulations on security that are appropriate for government systems managing confidential information should be applied to this educational system. There have been no moves to establishing a more effective governance body for the system.

Resolving the issues raised by these factors has required prolonged discussions and these have been major contributors to the delay in developing the system. Some remain unresolved and it's unclear what this will mean for the final system.

Meanwhile, more and more teachers have simply moved on to using other services that support learning on an individual or school basis. They devote time and effort in making these services work and will be unwilling to change when and if the Glow+ system is finally deployed.

So the ending of the story may not be so happy, after all.

## Thank you

I'd like to thank Professor Muffy Calder, the Chief Scientific Advisor to the Scottish Government for offering me the opportunity to be involved in this project. It was both fun and inspiring, although it didn't always seem like that at the time.

I'd also like to thank the other members of the ICT Excellence in Education team. We didn't agree about everything but think that we all ended up believing that what we proposed had the potential to enhance our childrens' learning.

## Ian Sommerville

I wrote my first computer program as a Physics student in 1970 and I've been involved with software and software engineering ever since then. I was a sort of academic in computer science for many years although some of my colleagues thought I was a bit suspicious as I insisted on talking to and working with people in industry. In the early 1990s, I started thinking about why software systems went wrong and decided that this wasn't really a technical problem. So, I've been interested in the interaction of technical and socio-technical issues since then.

I once wrote a book about software engineering, which has been updated a few times and which, in various editions and languages, has sold nearly a million copies.

I retired in 2014 and now do various software and systems things including re-learning how satisfying programming can be.

THE END